%% file: main.tex
\documentclass[conference]{IEEEtran}
\IEEEoverridecommandlockouts
\usepackage{cite}
\usepackage{amsmath,amssymb,amsfonts}
\usepackage{algorithmic}
\usepackage{graphicx}
\usepackage{textcomp}
\usepackage{xcolor}
\usepackage{enumitem}
\usepackage{graphicx}
\usepackage{hyperref}
\usepackage{caption}
\usepackage[all]{nowidow}
\usepackage{pgfplots}
\usepackage{multirow}
\usepackage{tablefootnote}
\usepackage{booktabs}
\usepackage[margin=1in]{geometry}
\pgfplotsset{compat=1.18}
\renewcommand\IEEEkeywordsname{Index Terms}

\graphicspath{{Images/}}
\begin{document}

\nocite{gitrepo}
\nocite{vectorStore}
\nocite{mistral}
\nocite{stackexchange}

\title{Repairing Responsive Layout Failures Using Retrieval Augmented Generation
}

\author{
\centering
\IEEEauthorblockN{Tasmia Zerin}
\IEEEauthorblockA{\textit{IIT, University of Dhaka} \\
Dhaka, Bangladesh \\
bsse1128@iit.du.ac.bd}
\and
\IEEEauthorblockN{Moumita Asad}
\IEEEauthorblockA{\textit{University of California, Irvine} \\
Irvine, California, USA \\
moumitaa@uci.edu}
\and
\IEEEauthorblockN{B M Mainul Hossain}
\IEEEauthorblockA{\textit{IIT, University of Dhaka} \\
Dhaka, Bangladesh \\
mainul@iit.du.ac.bd}
\and
\IEEEauthorblockN{Kazi Sakib}
\IEEEauthorblockA{\textit{IIT, University of Dhaka} \\
Dhaka, Bangladesh \\
sakib@iit.du.ac.bd}
}

\maketitle

\begin{abstract}
\input{abstract}
\end{abstract}
\begin{IEEEkeywords}
Responsive Web Design, Automated Program Repair, RAG-based tool, LLM, Stack Overflow
\end{IEEEkeywords}

\section{Introduction}
Designing a webpage responsive to all screens is challenging due to limited screen space, causing elements to overlap, overflow their containers, or extend beyond the visible area. These issues, known as Responsive Layout Failures (RLFs), are often difficult to detect and fix since they require extensive trial and error and modifications to multiple HTML elements and CSS properties \cite{automated_repair}. An automated solution that repairs layouts similar to human developers while preserving the original design would be beneficial.\\
\looseness=-1
In recent years, researchers have proposed techniques to detect, localize and repair RLFs. Walsh et al. \cite{automated_detection} first introduced a technique to automatically detect RLFs by comparing the positioning of elements relative to one another at different screen sizes. {\normalfont \textsc{LocaliCSS}} \cite{localicss} offers fine-grained localization by identifying the specific elements and CSS properties responsible for a detected RLF. To repair the detected layout issues, Layout DR \cite{automated_repair} creates hotfixes by modifying the CSS properties of the entire layout. However, existing approaches failed to combine localization with targeted repairs, causing inefficient repair solutions.\\
Recently, Large Language Models (LLMs) have shown promise in automated program repair \cite{llm4SE}. Hence they can be applied to repair RLFs as well, however, they often lack developer domain knowledge and in-built mechanism for developer validation or correction. On the other hand, popular developer Q\&A platforms such as Stack Overflow (SO) provide a vast space for knowledge sharing \cite{stack_obsolete}. SO allows developers to access solutions to problems faced by others and facilitate the exchange of expertise between developers.\\
In this paper, we propose ReDeFix (\underline{Re}sponsive \underline{De}sign \underline{Fix})\cite{gitrepo}, a Retrieval-Augmented Generation (RAG)-based approach for repairing RLFs. A knowledge base from SO discussions is constructed to enhance the quality of LLM-generated solutions. ReDeFix begins by localizing the faulty elements and CSS properties using {\normalfont \textsc{LocaliCSS}} \cite{localicss}. Next, it retrieves relevant SO discussions to help LLM understand how developers typically approach repairs. These discussions, along with the specific RLF context, are then combined into a prompt for LLM to generate an effective repair. This repair process is iterative, incorporating feedback after each attempt until a correct patch is produced.\\
Evaluation on 13 responsively designed webpages shows that ReDeFix successfully repaired 38 out of 43 RLFs, achieving an 88\% repair accuracy. To measure the impact on layout and aesthetics, a study with software engineers (SE) was performed. It revealed that, 85\% of the repaired layouts were preferred as correct, and 70\% were deemed aesthetically pleasing. This implies that the generated patches are effective and rarely degrade layout and aesthetics. Moreover, ReDeFix improved 11 out of 12 original incorrect layouts to correct states and enhanced aesthetics for 9 out of 11 non-aesthetic layouts. This confirms its ability to repair accurately without degrading layout quality.

\section{Literature Review}
Detecting different types of layout failures in web applications has been extensively explored in the literature, including \textit{Responsive Layout Failures} (RLFs). However, comparatively there has been less focus on automated approaches to repair them.\\
For responsively designed web pages, Walsh et al. \cite{automated_detection} first introduced a method to automatically detect RLFs using the \textit{Responsive Layout Graph} (RLG). It has HTML elements as nodes and their relationships as edges. This method compares RLGs of a page’s old and new versions using {\normalfont \textsc{ReDeCheck}} \cite{redecheck} to detect failures, but requires both versions.\\
Addressing this, to detect RLFs, Walsh et al. \cite{no_explicit_oracle} later presented another version of {\normalfont \textsc{ReDeCheck}} where the layout of a responsive webpage is compared against itself. It compares the relative positioning of elements at different screen sizes, also known as viewport widths. This study introduced five types of RLFs prevalent in real-world web pages. These types have been extensively used in the following works, including this paper. They are, \textit{Element Collision, Element Protrusion, Viewport Protrusion, Small-Range} and \textit{Wrapping Elements}. Although {\normalfont \textsc{ReDeCheck}} reliably detects these types of RLFs, it does not provide automatic repair of them.\\
For localizing the properties causing an RLF, Tasmia et al. \cite{localicss} proposed a method that heuristically searches for HTML elements and CSS properties potentially responsible to cause an RLF. The output of their approach is a ranked list, consisting of elements and their properties that have the most impact on an RLF.\\
To repair RLFs, Althomali et al. \cite{automated_repair} proposed a tool called {\normalfont \textsc{Layout DR}}. It sources layouts from either side of the affected viewport range, free from RLFs. The layouts are scaled and transformed within the failure range to create two \textit{“hotfixes”}. Since the entire layout is modified, it causes a change to almost every property value explicitly defined by developers.\\
On the other hand, recent advancements in automated program repair have benefited significantly from LLM-based approaches \cite{thinkrepair, repairagent}. While LLMs have improved code generation and repair tasks, these models are retrained infrequently, generate solutions non-deterministically, and often lack mechanisms for combining community-driven knowledge \cite{challengesLLM}. To overcome these limitations, recent studies have explored Retrieval-Augmented Generation (RAG), which leverages external knowledge sources such as Stack Overflow and other Q\&A platforms \cite{stackrag, ragfix, sosecure}. However, repairing RLFs using LLMs remains an unexplored area.
\looseness=-1

\section{Methodology}
\begin{figure*}[t]
    \centering
    \includegraphics[width=1\textwidth]{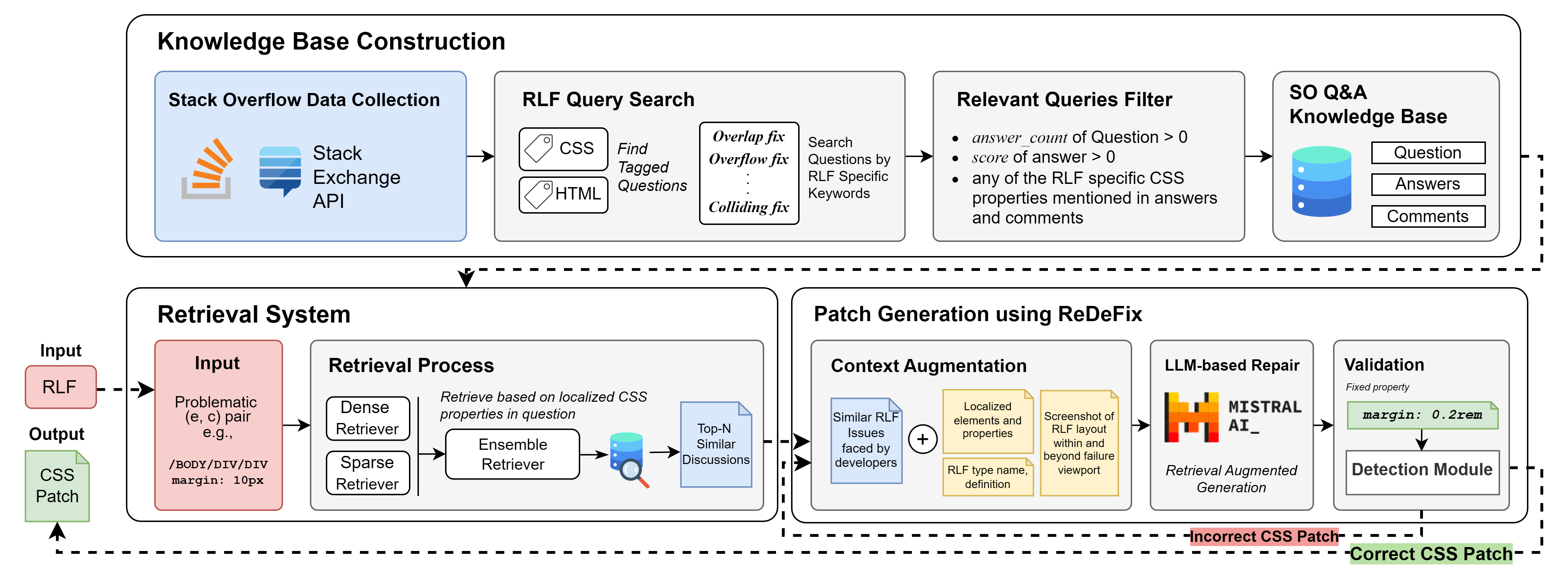}
    \caption{Overview of ReDeFix}
    \label{fig:redefix-overview}
\end{figure*}

In this study, we propose a novel RAG-based system to repair RLFs by using SO as a data source. As shown in Figure \ref{fig:redefix-overview}, our approach takes the problematic HTML elements and CSS properties as input. To address these RLFs, relevant SO queries are retrieved from the knowledge base to provide additional context on the repair styles of developers. Then, LLM is prompted to fix the failure along with this context.

\subsection{Knowledge Base Construction}
We utilized SO to extract relevant RLF-related questions, their answers and comments. SO is potentially useful to repair many failures as it contains millions of posts with years of accumulation \cite{miningstack}. We searched questions based on two tags, \texttt{CSS} and \texttt{HTML}, as majority of the RLF-related questions have one of these tags associated with them. To identify relevant RLF discussions, we used RAKE algorithm \cite{rake} to extract keywords from the definition of each RLF mentioned in \cite{localicss}. To search each RLF-specific SO questions, we chose the keywords that uniquely identify an RLF type, e.g., \textit{elements collide} for element collision, \textit{appear outside screen} for viewport protrusion. We then broaden the range of questions by adding similar keywords, such as \textit{elements overlap} alongside \textit{elements collide} as \textit{overlap} is a synonym for \textit{collide}. As SO has too many queries in various formats, we couldn't analyze them all. Therefore, we created this list to specifically identify a large, relevant set of questions for each RLF within the scope of our study.\\
For each RLF type, after applying these filters, we extracted the answers along with all their associated comments. Questions having no answers and comments were discarded. Similarly, answers having a SCORE of 0 or less than 0 were considered non-verified and thus discarded. Lastly, to find relevant answers and comments, we filtered the ones having mention of any of the RLF-specific properties. Since all collected questions are originally in HTML format, we performed standard data cleaning procedures. We removed all HTML tags except for {\small \texttt{<code>}} tags, which were preserved to maintain the integrity of code blocks within the posts.\\
These results are stored in the vector database as documents, where each RLF type contains a separate set of SO results. Each of the results contains the question's ID, LINK, TITLE and BODY in its metadata. Our knowledge base contains 334 extracted questions, with 522 answers and 1855 comments. This serves by capturing collective developers' feedback while they faced similar issues during the repair of RLFs.\\
\vspace{-1.0em}
\subsection{Retrieval System}
Given a list of problematic HTML elements and CSS properties, we retrieve discussions related to these properties from our knowledge base (Figure \ref{fig:redefix-overview}). Here we combine the problematic property names as a query to find repair discussions only for those properties. The query is used to retrieve the top five questions. Then with each question, its answers and associated comments are included as context and sent with the prompt.\\
\looseness=-1
We have used an Ensemble Retriever \cite{ensemble}, combining BM25 \cite{bm25} and VectorStore Retrievers \cite{vectorStore}. This ensembling enables to use both, dense (VectorStore) and sparse (BM25) retrievers. The sparse retriever is good at finding relevant documents based on keywords, while the dense retriever is good at finding relevant documents based on semantic similarity.  We selected BM25 because prior studies \cite{whybm25} have demonstrated its effectiveness for code-to-code retrieval. On the other hand, to find the properties in the discussion text, similarity search of VectorStore can be very effective.
\looseness=-1

\subsection{Patch Generation and Validation}
\begin{figure}[h]
    \centering
    \includegraphics[width=1\linewidth]{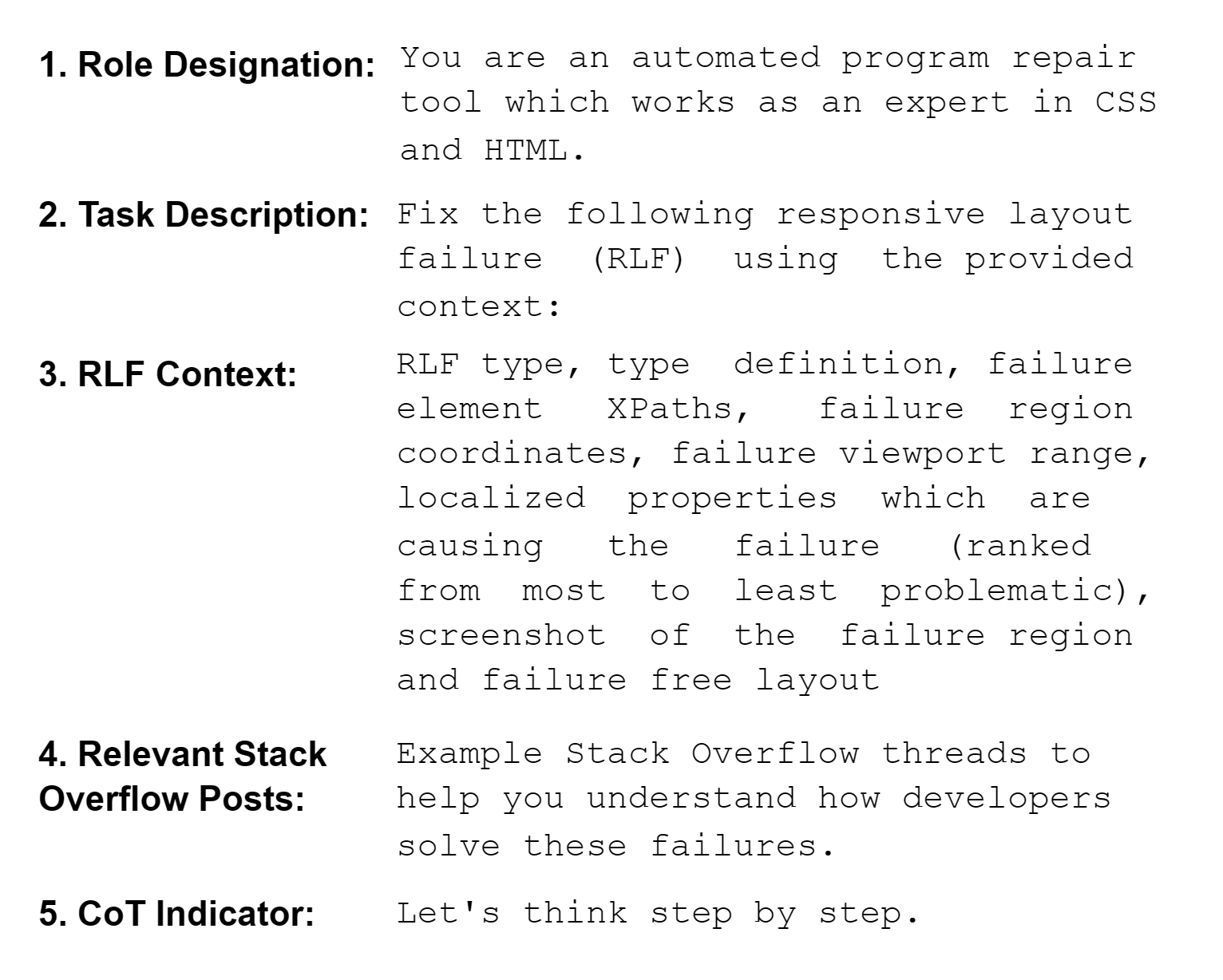}
    \caption{Example Prompt}
    \label{fig:Example Prompt}
\end{figure}

The main goal of this component is to augment the original prompt with similar issues collected from SO. The original prompt contains the context of an RLF, which includes the involved elements and properties, RLF type, its definition as stated in \cite{localicss}, the failure elements with their coordinates, and screenshots of the RLF segment inside and outside of the failure viewport range. These components are obtained by using the localization approach of {\normalfont \textsc{LocaliCSS}} \cite{localicss}. The target of this context is to help LLM understand the layout and generate a patch accordingly. The augmented context is then sent as a prompt to the LLM. The prompt used in our approach involves five important components as illustrated in Figure \ref{fig:Example Prompt}:

\noindent \textbf{- Role Designation:}
  Our approach starts with an instruction to assign LLM as an automated program repair tool \cite{better_zero_with_roleplay}. The assigned role provides context about the LLM’s identity and background.
\smallskip

\noindent
\textbf{- Task Description:}
  LLM is provided with the description as illustrated in Figure \ref{fig:Example Prompt}.
\smallskip

\noindent\textbf{- RLF Context:}
  Our approach provides LLM with the components working as context for the RLF.
\smallskip

\noindent \textbf{- Relevant SO Posts:}
  The retrieved SO posts are augmented with the original prompt to add external knowledge in the repair process.
\smallskip

\noindent \textbf{- Chain-of-Thought (CoT) Indicator:}
  LLM is instructed to think step-by-step to fix an RLF. Here, we follow the best practice to use CoT for enhancing the reasoning of LLM \cite{CoT} and adopt the same prompt named {\small“\texttt{Let’s think step by step}”}.
\smallskip

Following this, we can obtain the model output, which contains the thought process of LLM and the candidate patch created with the fixed properties of the problematic elements. These will be verified in the validation step as shown in Figure \ref{fig:redefix-overview}.\\
The generated patch is injected into the CSS of the target page by first creating a selector for each affected HTML element. To ensure these new rules override existing styles, every property in the patch is marked with {\small\texttt{!important}}. To restrict the patch to the failure range, the element and its properties are encapsulated within a media rule spanning the failure range. This patch is added to the webpage and then sent back to the detection module, which verifies that the original RLF has been eliminated without introducing any new layout failures. If both conditions are met, the patch is accepted. Otherwise, ReDeFix reconstructs the prompt and appends the failed patch to the end of the original prompt. Here, the failing information is added into a template: {\small“\texttt{The fixed version is still not correct-\{last generated patch\}. Please fix it again. Let’s think step by step.}”} Then, ReDeFix interacts with LLM using the new prompt to generate a new fixed solution. This iterative process continues until a correct patch is obtained or the prompt exceeds its maximum token limit.

\section{Experimental Setup and Result Analysis}
We designed an empirical study to evaluate the performance of the proposed ReDeFix by investigating two research questions (RQs):\\
\textbf{RQ1:} \textit{How accurate is ReDeFix in repairing responsive layout failures in webpages?}\\
\textbf{RQ2:} \textit{How do users perceive the quality of the repaired version generated by ReDeFix?}

\subsection{Subjects}
We selected our test subjects from the set of web pages previously used by {\normalfont \textsc{LocaliCSS}} \cite{localicss}. Their study localized 58 failures across 20 webpages; however, we excluded small-range failures since these require only simple media-query adjustments. Furthermore, the study of {\normalfont \textsc{LocaliCSS}} revealed that some failures were marked as No Problem (NP) by manual inspection. Additionally, some pages had a single RLF, which was later labelled as NP. Excluding above mentioned cases leaves us 13 webpages and a total of 43 RLFs for the evaluation.

\subsection{\textit{Answer to RQ1}}
\subsubsection{\textbf{Experimental Setup}} To evaluate the accuracy, we ran our approach on each of the 43 RLFs. Our automated validation phase gave us an output of the successful and failed repairs. Additionally, a manual inspection was performed. As one failure can have multiple ways to fix it, instead of having a manually annotated ground truth, the author added each generated patch to its corresponding webpage and inspected whether it fixes the RLF or not, without introducing any new RLFs.\\
We implemented our approach by using LangChain framework \cite{langchain_llm} with the API endpoint of Mistral Small 3.1 (24B) \cite{mistral}. It is a lightweight open-source LLM model with multimodal understanding and an expanded context window of up to 128k tokens. We used default values for all the hyper-parameters for our LLM model. For retrieval, we combined BM25 and semantic similarity search in an ensemble and experimentally evaluated various weight combinations to collect the most relevant discussions. This resulted in weights of 0.8 and 0.2, respectively. Finally, we retrieved SO posts via Stack Exchange API \cite{stackexchange} to access the up-to-date content.

\input{Tables/Result-Table-1}

\subsubsection{\textbf{Results}}
Our approach successfully repaired 38 out of 43 RLFs, as detailed in Table \ref{tab:Evaluation of ReDeFix and Zero-Shot Repair}. To evaluate the impact of augmenting the LLM with external knowledge, we compared the baseline LLM using zero-shot reasoning against the same LLM enhanced with additional context retrieved from our Stack Overflow knowledge base.\\
\looseness=-1
Table \ref{tab:Evaluation of ReDeFix and Zero-Shot Repair} shows promising results for RAG-based LLMs. Specifically, LLM's repair accuracy increased from 60.4\% in the zero-shot setting to 88.3\% using RAG in ReDeFix. Zero-shot patches frequently lacked advanced CSS properties and introduced unnecessary additional properties alongside the required changes. Incorporating developer knowledge from SO significantly improved the LLM's ability to adopt developer perspective and generate accurate repairs with essential properties only. One element protrusion RLF was repaired by adding {\small \texttt{box-sizing: border-box}}, but zero-shot patch was missing this property, thus unable to repair.\\
However, ReDeFix failed to repair 5 RLFs as seen in Table \ref{tab:Evaluation of ReDeFix and Zero-Shot Repair}. Here, for \textit{DjangoREST} RLF, our generated repair distorted the entire layout and introduced a new failure. In one of the three \textit{PepFeed} RLFs, it produced a hallucinated repair using a class from SO discussion. It failed to remove the original RLF in the other two RLFs and the RLF of \textit{MidwayMeetup}. Manual inspection revealed that these RLFs were not properly localized by {\normalfont \textsc{LocaliCSS}} \cite{localicss}, hence localized properties were absent. Our repair approach achieves 95\% accuracy when cases failing due to {\normalfont \textsc{LocaliCSS}} limitations are excluded. Hence, we can conclude that using a RAG with SO knowledge base enhances LLMs in code repair.

\subsection{\textit{Answer to RQ2}}
\subsubsection{\textbf{Experimental Setup}} 

To evaluate the quality of our proposed approach, we conducted a user study with five front-end software engineers (SE) having 1 to 5 years of professional experience.\\
Initially, we identified 43 RLFs across various elements and viewport ranges, with some RLFs appearing multiple times or within the same viewport range. For the user study, we combined such RLFs, and omitted the ones that we were unable to repair. This resulted in 20 unique pairs of layout screenshots to assess in our survey. Each pair featured the original (before repair) and repaired (after repair) view, presented side-by-side in the survey. To ensure unbiased responses, we labeled the screenshots simply as version 1 and 2. Additionally, a reference image was provided for each of the layout. It is an ideal layout of the corresponding webpage free from RLFs. We captured screenshots at the lowest viewport size of the RLF range using the browser’s inspect tool. Since in this range, the lowest viewport size contains the most constrained, worst-case rendering of the layout.\\ 
We asked each SE to select (1) Is the layout correct? (2) Is the layout aesthetically pleasing?. We briefed them on the definition of a layout as ``correct" if (1) it is positioned correctly (e.g., no visible RLFs, correctly aligned), (2) it is not deviated from the reference layout. Similarly, a layout as ``aesthetic" if it is visually appealing (e.g., good spacing and margins).
\vspace{0.2em}
\subsubsection{\textbf{Results}}
The user study results, depicted in Table \ref{tab:repair score table} illustrate the satisfaction ratings for the 20 layouts. For considering a layout as \textit{correct} and \textit{aesthetic}, we counted the maximum vote for both. In case of aesthetic, one layout had same vote for multiple decisions. It was resolved by an additional vote by the first author. Finally, the study revealed that before repair, 8 layouts or 40\% were considered as correct based on the maximum vote. After repair, this increased to 17 or 85\% layouts as correct. Similarly for aesthetics, before repair there were 9 or 45\% aesthetic layouts, which increased to 14 or 70\% after repair.
\looseness=-1
\input{Tables/result-2-whole}
\\
For a more rigorous analysis, we employed the Wilcoxon Signed-Rank test \cite{wilcoxon} separately on the set of total response votes for correctness and aesthetics of each layout and measured the difference level between before and after repair. The test yielded a p-value of approximately $p \approx 0.026 < 0.05$ for correct layout, which indicates a significant difference in before and after repair. Regarding aesthetics, although no statistical significance was found at $p < 0.05$, we observed a slight numerical improvement in the proportion of layouts rated as aesthetic after repair. As opinions on aesthetics are subjective, improvements tend to be smaller, and people’s opinions vary more.\\
However, we investigated the layouts where the repaired version was not preferred by the SEs. In one example, two divs were overlapping on one another, but was not visible. When the overlap was fixed, a small gap appeared between them, which made the original layout and aesthetics better. On the other hand, engineers marked one layout as incorrect both before and after repair, and two layouts as not aesthetic in either version. Such layouts had incorrect placements before repair, which got fixed, but then introduced a different problem. Some of the other votes varied from person to person due to the subjective nature of aesthetic preferences. Overall, our approach demonstrates its quality clearly through both statistical and descriptive analyses.

\subsection{Threats to Validity}
Generalizability of webpages used in our study is one of the threats to validity. Hence, we selected webpages of varying sizes, as shown in Table \ref{tab:Evaluation of ReDeFix and Zero-Shot Repair}. This approach is consistent with prior studies \cite{automated_repair, localicss}. Another possible threat comes from {\normalfont \textsc{LocaliCSS}} \cite{localicss}. Since we rely on this for our input, its accuracy directly influences the results. LLMs typically generate non-deterministic results, causing a potential threat to validity. To mitigate this, we ran the model five times and took the most frequently generated patch. Moreover, as LLMs have limitations of max tokens, we restricted our input to the top five SO discussions to prevent overflow of token limits. Finally, subjectivity remains a potential threat when manually validating patches. Hence, we took opinions from five expert SEs to check the effectiveness.
\looseness=-1

\section{Conclusion and Future Work}
In this work, we present ReDeFix, an LLM-based RLF repair approach using RAG to leverage Stack Overflow (SO) discussions. The generated patch is then validated to ensure the RLF is repaired and no new RLFs exist. Evaluations show that it successfully repaired 88\% of the existing 43 RLFs. A study on software engineers resulted in preference for 85\% of the repaired layouts as correct, and 70\% as aesthetic. Overall, these show the effectiveness of ReDeFix in repairing RLFs without distorting the layouts and aesthetics. Future research can be directed to understand layout aesthetics during repair to preserve the visual appeal of a webpage.

\section{Acknowledgements}
This research is supported by the Fellowship from ICT Division, Government of Bangladesh; No-56.00.0000.000.052.33.0002.24-97, dated 22.05.2025.

\clearpage
\bibliographystyle{IEEEtran}
\input{main.bbl}

\end{document}

%% file: abstract.tex
Responsive websites frequently experience distorted layouts at specific screen sizes, called Responsive Layout Failures (RLFs). Manually repairing these RLFs involves tedious trial-and-error adjustments of HTML elements and CSS properties. In this study, an automated repair approach, leveraging LLM combined with domain-specific knowledge is proposed. The approach is named ReDeFix, a Retrieval-Augmented Generation (RAG)-based solution that utilizes Stack Overflow (SO) discussions to guide LLM on CSS repairs. By augmenting relevant SO knowledge with RLF-specific contexts, ReDeFix creates a prompt that is sent to the LLM to generate CSS patches. Evaluation demonstrates that our approach achieves an 88\% accuracy in repairing RLFs. Furthermore, a study from software engineers reveals that generated repairs produce visually correct layouts while maintaining aesthetics. 


%% file: Tables/Result-Table-1.tex
\begin{table}[]
\caption{Evaluation of ReDeFix on different settings}
\label{tab:Evaluation of ReDeFix and Zero-Shot Repair}
\resizebox{\columnwidth}{!}{%
\begin{tabular}{lrrr}
\hline
\textbf{Subject Name} & \multicolumn{1}{c}{\textbf{Total RLFs}} & \multicolumn{1}{c}{\textbf{\begin{tabular}[c]{@{}c@{}}Zero-Shot\\ Repair\end{tabular}}} & \multicolumn{1}{c}{\textbf{\begin{tabular}[c]{@{}c@{}}ReDeFix\\Repair\end{tabular}}} \\ \hline
3MinuteJournal & 4 & 4 & \textbf{4} \\
Ardour & 2 & 2 & \textbf{2} \\
Bower & 3 & 3 & \textbf{3} \\
BugMeNot & 3 & 2 & \textbf{3} \\
Django & 3 & 1 & \textbf{3} \\
DjangoREST & 1 & 0 & \textbf{0} \\
HotelWiFiTest & 1 & 1 & \textbf{1} \\
MantisBT & 6 & 3 & \textbf{6} \\
MidwayMeetup & 1 & 0 & \textbf{0} \\
PepFeed & 5 & 1 & \textbf{2} \\
PDFescape & 2 & 2 & \textbf{2} \\
Selenium & 6 & 4 & \textbf{6} \\
WillMyPhoneWork & 6 & 3 & \textbf{6} \\ \hline
\textbf{Total} & \textbf{43} & \textbf{\begin{tabular}[c]{@{}r@{}}26\\(60.4\%)\end{tabular}} & \textbf{\begin{tabular}[c]{@{}r@{}}38\\(88.3\%)\end{tabular}} \\ \hline
\end{tabular}%
}
\vspace{-1.5em}
\end{table}

%% file: Tables/result-2-whole.tex
\renewcommand{\arraystretch}{1.3}
\begin{table}[ht]
\caption{ Score Comparisons: Before vs After Repair}
\resizebox{\columnwidth}{!}{%
\begin{tabular}{cccc|cccc}
\multicolumn{4}{c|}{Layout Score} & \multicolumn{4}{c}{Aesthetics Score} \\ \hline
 & \multicolumn{2}{c}{After Repair} &  &  & \multicolumn{2}{c}{After Repair} &  \\ \cline{2-3} \cline{6-7}
\begin{tabular}[c]{@{}c@{}}Before \\ Repair\end{tabular} & L & $L^0$ & Total & \begin{tabular}[c]{@{}c@{}}Before \\ Repair\end{tabular} & A & $A^0$ & Total \\ \hline
L & 6 & 2 & 8 & A & 5 & 4 & 9 \\
$L^0$ & 11 & 1 & 12 & $A^0$ & 9 & 2 & 11 \\ \hline
Total & 17 & 3 & 20 & Total & 14 & 6 & 20 \\ \hline
\\
\end{tabular}%
}
 \begin{minipage}{\columnwidth}
\footnotesize
\textit{L=Correct Layout $|$ $L^0$=Not Correct Layout \\ A=Aesthetic $|$ $A^0$=Not Aesthetic}
\end{minipage}
\label{tab:repair score table}
\vspace{-0.8em}
\end{table}